\documentstyle[12pt]{article}
\newcommand{\ltwid}{\raise.3ex\hbox{$<$\kern-.75em\lower1ex\hbox{$\sim
$}}}

\title{{\hfill\normalsize NSF-ITP-94-44}\\[1.0cm]
Catalysis of Dynamical Flavor Symmetry Breaking
 by a Magnetic Field in $2+1$ Dimensions}

\author{{\sl V.P. Gusynin $^1$, V.A. Miransky $^{1,2}$, and I.A.
Shovkovy $^1$}\\
{\sl $^1$ Bogolyubov Institute for Theoretical Physics,
252143 Kiev, Ukraine}\\
{\sl $^2$ Institute for Theoretical Physics,
University of California,}\\
{\sl Santa Barbara, CA 93106-4030}}
\date{}
\begin{document}
\maketitle

\vfill

\begin{abstract}
It is shown that in $2+1$ dimensions, a constant magnetic field is a strong
catalyst of dynamical flavor symmetry breaking, leading to generating a fermion
dynamical mass even at the weakest attractive interaction between fermions. The
effect is illustrated in the Nambu-Jona-Lasinio model in a magnetic field. The
low-energy effective action in this model is derived and the thermodynamic
properties of the model are established. The relevance of this effect for
planar condensed matter systems is pointed out.
\end{abstract}

\vfill
\eject

Relativistic field models in $2+1$ dimensional space-time have been a subject
of considerable interest: Their sophisticated dynamics is interesting in
itself; they also serve as effective theories for the description of long
wavelength excitations in planar condensed matter systems [1,2]; also,
their dynamics imitates the dynamics of $3+1$ dimensional theories at high
temperature.

In this Letter, we will show that a constant magnetic field acts as a
strong catalyst of dynamical flavor symmetry breaking  in $2+1$ dimensions,
leading to generating a fermion dynamical mass even at the
weakest attractive interaction between fermions.
We would like to stress that this effect is universal, {\em i.e.,\/}
model independent, in $2+1$ dimensions. This point  may be important
in connection with considering this effect in such condensed
matter phenomena as the quantum Hall
effect [1] and high temperature superconductivity [2], and in
$3+1$ dimensional theories at high temperature.

Let us begin by considering the basic points in the problem of a relativistic
fermion in
a constant magnetic field $B$ in $(2+1)$-dimensions. The Lagrangian density
is
\begin{equation}
{\cal L} = \frac{1}{2} \left[\bar{\Psi},(i\gamma^\mu D_\mu-m)\Psi\right],
\label{eq:1}
\end{equation}

where the covariant derivative $D_\mu$ is
\begin{equation}
D_\mu=\partial_\mu-ie A^{ext}_\mu,\qquad A^{ext}_\mu =-B x_2 \delta_{\mu1}.
\label{eq:2}
\end{equation}
We use four-component spinors corresponding to a reducible representation of
the Dirac algebra [3]:
\begin{equation}
\gamma^0=\left( \begin{array}{cc} \sigma_3&0\\ 0&-\sigma_3\end{array}\right),
  \gamma^1= \left( \begin{array}{cc} i\sigma_1 & 0\\ 0 & -i\sigma_1\end{array}
\right),
\gamma^2= \left( \begin{array}{cc} i\sigma_2 & 0\\ 0 &- i\sigma_2 \end{array}
\right).
\end{equation}
At $m=0$, the Lagrangian (\ref{eq:1}) is
invariant under the flavor $U(2)$ transformations with generators $T_0=I,
T_1=\gamma^5,T_2=\frac{1}{i}\gamma^3,T_3=\gamma^3\gamma^5$ where
$\gamma^5=i\gamma^0\gamma^1\gamma^2\gamma^3$. The mass term breaks $U(2)$ down
to $U(1)\times U(1)$ with generators $T_0$ and $T_3$.

Our  basic observation is that in $2+1$ dimensions,  as $m\to0$,
{\em spontaneous\/} $U(2)$ symmetry breaking takes place for fermions in a
constant magnetic field. In order to prove this, we will show that in the limit
$m\to0$, the flavor condensate $\langle0|\bar{\Psi}\Psi|0\rangle$ is nonzero:
$\langle0|\bar{\Psi}\Psi|0\rangle=-\frac{|eB|}{2\pi}$.

The condensate is expressed through the fermion propagator
$S(x,y)=\langle0|T(\Psi(x)\bar{\Psi}(y))|0\rangle$.
The propagator $S$ can be calculated by using the Schwinger (proper time)
approach [4]. It is
\begin{equation}
S(x,y) = \exp(ie\int^x_y A_\lambda^{ext} dz^\lambda) \tilde{S} (x-y),
\label{eq:4}
\end{equation}
where the integral is calculated along the straight line, and the Fourier
transform of
$\tilde{S}(x)$ (in Euclidean space) is

\begin{eqnarray}
\tilde{S}(k) &=& -i \int^\infty_0ds \exp\left[-s\left(m^2+k^2_3+ {\bf k}^2
\frac{\tanh(eBs)}{eB s}\right) \right]
\Big(- k_\mu \gamma_\mu +m+ \nonumber\\
&+& \frac{1}{i} (k_2\gamma_1-k_1\gamma_2) \tanh (eBs) \Big)
\left(1+\frac{1}{i} \gamma_1\gamma_2\tanh (eB s)\right)
\end{eqnarray}
($k_3=-ik^0$ and $\gamma_\mu$ are antihermitian matrices). The condensate is

\begin{eqnarray}
\langle0|\bar{\Psi}\Psi|0\rangle &=& - \lim_{x \to y} tr S(x,y) = -
\frac{i}{(2\pi)^3} tr \int d^3 k
\tilde{S} (k) = - \lim_{\Lambda\to\infty}
\frac{4m}{(2\pi)^3}\int d^3k\int^\infty_{1/\Lambda^2}ds  \label{eq:6}\\
&\cdot&\exp\left[-s\left(m^2+k_3^2+{\bf k}^2 \frac{\tanh(eBs)}{eBs}
\right) \right] \overrightarrow{m\to0}\  - \frac{|eB|}{2\pi}
 \nonumber
\end{eqnarray}
($\Lambda$ is an ultraviolet cutoff).
Thus in a constant magnetic field, spontaneous breakdown of the $U(2)$
symmetry takes place even though fermions do not acquire a dynamical
mass. We note that in $3+1$ dimensions, the result would be
$\langle0|\bar{\Psi}\Psi|0\rangle\sim
 m\ln m\longrightarrow0$ as $m\to0$.
Therefore, this
is a specific $2+1$ dimensional phenomenon.

What is the physical basis of this
phenomenon? In order to answer this question, we note that the singular,
$\frac{1}{m}$, behavior of the integral in Eq.~(\ref{eq:6}) is formed
at large, $s\to\infty$, distances ($s$ is the proper time coordinate).
Actually, one can see from Eq.~(\ref{eq:6}) that the magnetic field effectively
removes
the two space dimensions in the infrared region, thus reducing the dynamics to
a
one-dimensional dynamics which has much more severe infrared singularities.
{}From this viewpoint, the action of the magnetic field in this problem is
similar to that of the Fermi surface in the BCS theory [5].

This point is intimately connected with the form of the energy spectrum of
fermions in a constant magnetic field. In $(2+1)$-dimensions, it is [6]:
\begin{equation}
E_n=\pm\sqrt{m^2+2|eB|n},\quad n=0,1,\dots \label{eq:7}
\end{equation}
(the Landau levels). Each Landau level is degenerate. At the lowest level with
$n=0$, the density of states is $\frac{|eB|}{2\pi}$; at the levels with
$n\geq1$, it is $\frac{|eB|}{\pi}$. As $m\to0$, the energy $E_0$ goes to
zero and therefore there is the infinite vacuum degeneracy in this case.

By using the explicit solution for the field $\Psi$ in this problem [6], we
find the charge operators connected with the broken symmetry:
\begin{eqnarray}
Q_1&=&\frac{1}{2}\int d^2x[\Psi^\dagger(x),T_1\Psi(x)]=i\sum\limits_p(a_{0p}^
\dagger d_{0-p}^\dagger-d_{0-p}a_{0p})\nonumber\\
&&+i\sum\limits_{n=1}^\infty\sum\limits_p
\left[(a_{np}^\dagger c_{np}-c_{np}^\dagger a_{np})+(b_{np}^\dagger d_{np}-
d_{np}^\dagger b_{np})\right],\nonumber\\
Q_2&=&\frac{1}{2}\int d^2x[\Psi^\dagger(x),T_2\Psi(x)]=\sum\limits_p(a_{0p}^
\dagger d_{0-p}^\dagger+d_{0-p}a_{0p})\nonumber\\
&&+\sum\limits_{n=1}^\infty\sum\limits_p
\left[(a_{np}^\dagger c_{np}+c_{np}^\dagger a_{np})+(b_{np}^\dagger d_{np}
+d_{np}^\dagger b_{np})\right].  \nonumber
\end{eqnarray}

Here $a_{np},c_{np}\,(b_{np},d_{np})$ are annihilation operators of fermions
(antifermions) from the $n$-th Landau level (there are two types of fermions
(antifermions) for the four-component fielt $\Psi$); the momentum $k={2\pi p/
L_1}\,(p=0,\pm 1,\ldots)$, $L_1$ is the size in the $x_1$-direction. A set of
the degenerate vacua $\vert\theta_1,\theta_2\rangle$ is $\vert\theta_1,
\theta_2\rangle\equiv\exp(iQ_1\theta_1+iQ_2\theta_2)\vert 0\rangle$ where
$\vert 0\rangle$ satisfies the condition $a_{np}\vert 0\rangle=b_{np}\vert0
\rangle=c_{np}\vert0\rangle=d_{np}\vert0\rangle=0$. As $L_1\rightarrow\infty$,
the vacua with different $\{\theta_1,\theta_2\}$ become orthogonal and define
(as usual in the case of spontaneous symmetry breaking) nonequivalent
representations of canonical commutation relations.

In each Fock space defined by a vacuum $\vert\theta_1,\theta_2\rangle$, there
are a
lot of "excitations" with nonzero momentum and zero energy created by the
operators $a_{0p}^\dagger,d_{0p}^\dagger$.
However, there are no genuine (i.e. with a nontrivial dispersion relation)
Nambu-Goldstone (NG) modes: At the lowest Landau level, the energy $E=0$
(since the Lorentz symmetry is broken by a magnetic field in this problem,
there is no contradiction with Goldstone's theorem). We shall show below that
even the weakest attractive interaction between fermions and antifermions is
enough to "resurrect" these modes (and a dynamical mass for fermions).

Let us consider the $2+1$ dimensional Nambu-Jona-Lasinio (NJL) model with
the $U(2)$ flavor symmetry:
\begin{equation}
{\cal L} = \frac{1}{2} \left[\bar{\Psi}, (i\gamma^\mu D_\mu)\Psi\right] +
\frac{G}{2} \left[ (\bar{\Psi}\Psi)^2+(\bar{\Psi}i\gamma^5\Psi)^2 +
(\bar{\Psi}\gamma^3\Psi)^2\right], \label{eq:10}
\end{equation}
where $D_\mu$ is the covariant derivative  (\ref{eq:2})  and fermion fields
carry an
additional, ``color", index $\alpha=1,2,\dots,N_c$. This theory is equivalent
to a theory with the Lagrangian density
\begin{eqnarray}
{\cal L}& =& \frac{1}{2} \left[\bar{\Psi}, \left( i\gamma^\mu D_\mu\right)
 \Psi\right] - \bar{\Psi}(\sigma+\gamma^3\tau+i\gamma^5\pi)\Psi \nonumber\\
&-& \frac{1}{2G} \left(\sigma^2+\pi^2+\tau^2\right)     \label{eq:11}
\end{eqnarray}
$\Big(\sigma=- G(\bar{\Psi}\Psi), \tau=-G(\bar{\Psi}\gamma^3\Psi),\ \pi =-G
(\bar{\Psi}i\gamma^5\Psi)\Big)$.  The effective action for these composite
fields is:

\begin{equation}
\Gamma(\sigma,\tau, \pi) = -\frac{1}{2G}\int d^3x(\sigma^2+\tau^2+\pi^2) +
\tilde{\Gamma}(\sigma,\tau,\pi), \label{eq:12}
\end{equation}
where
\begin{equation}
 \tilde{\Gamma}(\sigma,\tau,\pi) =- i Tr Ln \left[i\gamma^\mu D_\mu
- (\sigma+\gamma^3\tau+i\gamma^5\pi)\right].    \label{eq:13}
\end{equation}
As $N_c\to\infty$, the path integral over the composite fields is dominated by
stationary points of their action:
$\frac{\delta\Gamma}{\delta\sigma}=\frac{\delta\Gamma}{\delta\tau}=
\frac{\delta\Gamma}{\delta\pi}=0$. We will analyze the dynamics in this limit
by using the expansion of the action $\Gamma$ in powers of derivatives of the
composite fields.

We begin by calculating the effective potential $V$. Since $V$ depends only on
the $U(2)$-invariant $\rho^2=\sigma^2+\tau^2+\pi^2$, it is sufficient to
consider a configuration with $\tau=\pi=0$ and $\sigma$ independent of $x$.
Then, using the proper-time method [4], we find the potential from
Eqs.~(\ref{eq:4}), (\ref{eq:12}), and (\ref{eq:13}):
\begin{equation}
V(\sigma)=\frac{\sigma^2}{2G}+\tilde{V}(\sigma)=\frac{\sigma^2}{2G} +
\frac{N_c}{4\pi^{3/2}} \int^\infty_{1/\Lambda^2} \frac{ds}{s^{5/2}}
e^{-s\sigma^2} eB s \coth(eB s). \label{eq:14}
\end{equation}
 The gap equation,
$\frac{dV}{d\sigma}=0$, is
\begin{equation}
2\Lambda l \left(\frac{1}{g}-\frac{1}{g_c}\right) \sigma =
 \frac{\sigma}{|\sigma|l}
+\sqrt{2}\sigma\zeta \left(\frac{1}{2},1+\frac{(\sigma l)^2}{2}\right) + O
(1/\Lambda),      \label{eq:15}
\end{equation}
where the dimensionless coupling constant $g\equiv N_c \frac{\Lambda}{\pi}G$,
$g_c=\sqrt{\pi}$, $l\equiv1/|eB|^{1/2}$ is
 the magnetic length and $\zeta$ is the
generalized Riemann zeta function [7].  As $B\to0(l\to\infty)$, we recover the
known gap
equation [8]
\begin{equation}
\sigma|\sigma|=\sigma\Lambda \left(\frac{1}{g_c}-\frac{1}{g}\right)
  \label{eq:16}
\end{equation}
which admits a nontrivial solution only if $g$ is supercritical,
$g>g_c=\sqrt{\pi}$ (as Eq.~(\ref{eq:11}) implies, a solution to
Eq.~(\ref{eq:15}),
$\sigma=\bar{\sigma}$, coincides with the fermion dynamical mass,
$\bar{\sigma}=m_{dyn}$, and the dispersion relation for fermions is
Eq.(\ref{eq:7})
with $m$ replaced by $\bar\sigma$).
 We will show that a magnetic field changes the situation
dramatically: At $B\neq0$, a nontrivial solution exists at all $g>0$.

We shall first consider the case of subcritical $g,g<g_c$, which in turn can be
divided into two subcases: {\em a)} $g\ll g_c$ and {\em b)} $g\to g_c-0$
 (nearcritical $g$).  Assuming that $|\bar{\sigma} l|\ll1$ at $g\ll g_c$, we
find from
Eq.~(\ref{eq:15}):
\begin{equation}
m_{dyn}\equiv\bar{\sigma}\simeq \frac{|eB|g\sqrt{\pi}}{2\Lambda(g_c-g)}.
\label{eq:17}
\end{equation}
Since this equation implies that the condition
$|\bar{\sigma} l|\ll 1$ fulfills at all $g$
satisfying $(g_c-g)\gg\frac{|eB|^{1/2}}{\Lambda}$, the relation
(\ref{eq:17}) is
actually valid in that whole region.

At $g_c-g\ltwid\frac{|eB|^{1/2}}{\Lambda}$, introducing the scale
$m^*\equiv2\Lambda(\frac{1}{g}-\frac{1}{g_c})$, we find the equation
\begin{equation}
m^*l=\frac{1}{|\sigma|l}+\sqrt{2}\zeta
\left(\frac{1}{2}, \frac{(\sigma l)^2}{2}+1
\right)
\end{equation}
which implies that in the nearcritical region $m_{dyn}$ is
\begin{equation}
m_{dyn}=\bar{\sigma}\sim|eB|^{1/2}.  \label{eq:19}
\end{equation}
Thus in the scaling region, with $g_c-g\ltwid|eB|^{1/2}/\Lambda$, the
cutoff disappears from the observable $m_{dyn}$. This agrees with the
well-known fact that the critical value $g_c=\sqrt{\pi}$ is an ultraviolet
stable
fixed point at leading order in $1/N_c$ [8]. The relation (\ref{eq:19}) can be
considered as a scaling law in the scaling region.

At $g>g_c$, the analytic expression for $m_{dyn}$
 can be obtained at weak $|eB|$,
satisfying the condition $\frac{|eB|^{1/2}}{m_{dyn}^{(0)}}\ll1$, where
$m^{(0)}_{dyn}$ is the solution of the gap equation (\ref{eq:16}) with
$B=0$. Then,
using the asymptotic formula $\zeta(z,q)\longrightarrow$
 $\frac{1}{(z-1)q^{z-1}} \left[1+\frac{z-1}{2q}+\dots\right]$ as
$q\to\infty$ [7], we find
\begin{equation}
m_{dyn}=\bar{\sigma}=m_{dyn}^{(0)}
\left[1+\frac{(eB)^2}{12(m^{(0)}_{dyn})^4}\right], \label{eq:20}
\end{equation}
{\em i.e.,\/} $m_{dyn}$ increases with $B$ [9].

Let us now turn to calculating the kinetic term ${\cal L}_k$ in the action. The
$U(2)$ symmetry implies that the general form of ${\cal L}_k$ is
\begin{equation}
{\cal L}_k= N_c \frac{F^{\mu\nu}_1}{2} (\partial_\mu\rho_j\partial_\nu\rho_j) +
N_c
\frac{F^{\mu\nu}_2}{\rho^2} (\rho_j\partial_\mu\rho_j)
(\rho_i\partial_\nu\rho_i),    \label{eq:21}
\end{equation}
where ${\mbox{\boldmath$\rho$}}=(\sigma,\tau,\pi)$ and
$F^{\mu\nu}_1,F^{\mu\nu}_2$ are
functions of $\rho^2=\sigma^2+\tau^2+\pi^2$. To find the functions
$F^{\mu\nu}_1, F^{\mu\nu}_2$, one can use different methods. We used the method
of Ref.~[10]. The result is: $F^{\mu\nu}_1=g^{\mu\nu}F^{\mu\mu}_1$,
$F^{\mu\nu}_2=g^{\mu\nu}F^{\mu\mu}_2$ with

\begin{eqnarray}
F^{00}_1 &=& \frac{l}{8\pi} \left(\frac{1}{\sqrt{2}} \zeta \left(\frac{3}{2},
\frac{(\rho l)^2}{2}+1\right)+(\rho l)^{-3}\right), \nonumber\\
F^{11}_1 &=& F^{22}_1=\frac{1}{4\pi\rho}, \label{eq:22}\\
F^{00}_2 &=& - \frac{l}{16\pi} \left(\frac{(\rho l)^2}{2\sqrt{2}} \zeta
\left(\frac{5}{2}, \frac{(\rho l)^2}{2}+1\right) + (\rho l)^{-3} \right),
\nonumber\\
F^{11}_2 &=& F^{22}_2 = \frac{l}{8\pi}
\Bigg[\frac{(\rho l)^4}{\sqrt{2}} \zeta \left(\frac{3}{2}, \frac{(\rho l)^2}{2}
+1\right) +\sqrt{2}(\rho l)^2 \zeta \left(\frac{1}{2}, \frac{(\rho l)^2}{2}+1
\right)  \nonumber\\
&+& 2\rho l - (\rho l)^{-1}\Bigg].\nonumber
\end{eqnarray}
By using the asymptotic formulae for zeta functions [7], we find the following
spectrum for the excitations
$\sigma,\tau$, and $\pi$ from Eqs.~(\ref{eq:14}), (\ref{eq:21}) and
(\ref{eq:22}):

{\em a)} Subcritical, $g<g_c$,
 region:

 At $g_c-g\gg\frac{|eB|^{1/2}}{\Lambda}$ (where
$|\bar{\sigma}l|\ll 1,$ see Eq.~(\ref{eq:17})) we find:
\begin{eqnarray}
E_{\tau,\pi} &\simeq& \sqrt{2}(\bar{\sigma}l) ({\bf k}^2)^{1/2}
=\frac{g|eB|^{1/2}}{\sqrt{2}\Lambda(1-\frac{g}{g_c})} ({\bf k}^2)^{1/2}, \\
M^2_\sigma &\simeq& \frac{8\sqrt{2}(1-\frac{g}{g_c})}{\zeta(\frac{3}{2})g}
 \Lambda |eB|^{1/2}
\end{eqnarray}
(see Eq.~(\ref{eq:17})). Thus $\tau$ and $\pi$ are gapless NG modes. As the
interaction is switched off, $g\to0$,
 their energy becomes identically zero, and we
return  to the dynamics with spontaneous flavor symmetry breaking but without
genuine NG modes. Also, as
$g\to0$, the $\sigma$-mode decouples $(M_\sigma\to\infty)$.

In the nearcritical region, with $g_c-g\ltwid\frac{|eB|^{1/2}}{\Lambda}$, we
find from Eqs.~(\ref{eq:21}), (\ref{eq:22}):
\begin{equation}
E_{\tau,\pi}=f(\bar{\sigma}l)({\bf k}^2)^{1/2}
\end{equation}
where
$$
f(\bar{\sigma}l)=\left(\frac{2}{\bar{\sigma}l}\right)^{1/2}
\left(\frac{1}{\sqrt{2}} \zeta \left(\frac{3}{2}, \frac{(\bar{\sigma}l)^2}{2}
+1\right)+
(\bar{\sigma}l)^{-3}\right)^{-1/2}.
$$
Since in the nearcritical (scaling) region the parameter $\bar{\sigma}$ is
$\bar{\sigma}\sim|eB|^{1/2}=l^{-1}$, we see that the cutoff $\Lambda$
disappears from the observables $E_\tau$ and $E_\pi$ in the scaling region.

{\em b)} Supercritical $g$, $g>g_c$, and weak $B(l\to\infty)$:
\begin{eqnarray}
E_{\tau,\pi} &=& \left(1-\frac{1}{8(\bar{\sigma}l)^4}\right)
({\bf k}^2)^{1/2}, \\
M^2_\sigma &=& 6 \bar{\sigma}^2\left(1-\frac{3}{4}\frac{1}{(\bar{\sigma}l)^2}
\right),
\end{eqnarray}
where $\bar{\sigma}$ is given in Eq.~(\ref{eq:20}). These relations show that
the
magnetic field leads to decreasing both the velocity of the NG modes (it
becomes less than 1) and the mass (energy gap) of the $\sigma$ mode.

As is known, in this model, an interacting continuum $(\Lambda\to\infty)$
theory appears only at the critical value $g=g_c$ (the continuum theory is
trivial at $g<g_c$) [8]. At $B=0$, the continuum theory is in the symmetric
phase at $g\rightarrow g_c-0$ and in the broken phase at $g\rightarrow g_c+0$.
On the other hand, as follows from our analysis, in a magnetic field, it is in
the broken phase both at $g\rightarrow g_c-0$ and $g\rightarrow g_c+0$
(though the dispersion relations for fermions and collective excitations
${\mbox{\boldmath$\rho$}}$ are different at $g\rightarrow g_c-0$ and
$g\rightarrow g_c+0$).

The present model illustrates the general phenomenon in $2+1$ dimensions:
In the infrared
region, an external magnetic field reduces the dynamics of fermion pairing to
one-dimensional
dynamics (at the lowest Landau level) thus catalysing the generation of a
dynamical mass for
fermions. A concrete sample of dynamical symmetry breaking is of course
different in different models.

We also studied the thermodynamic properties of the NJL model
(\ref{eq:10}) at finite
temperature $T$ and finite fermion density $n$. In leading order in $1/N_c$, we
found that the effective (thermodynamic) potential is:
\begin{eqnarray}
V_{\beta,\mu}(\sigma) &=& \frac{\sigma^2}{2G} + \frac{N_c}{4\pi^{3/2}l^3}
\int^\infty_0 \frac{dt}{t^{3/2}} e^{-(tl^2\sigma^2)} \coth t\nonumber\\
&\cdot& \Theta_4 \left( \frac{i}{2}\mu\beta|
\frac{i}{4\pi t}(\frac{\beta}{l})^2
\right),                         \label{eq:28}
\end{eqnarray}
where $\beta=1/T$, $\mu$ is a chemical potential and $\Theta_4$ is the fourth
Jacobian theta function [7]. The analysis of Eq.~(\ref{eq:28})
 (which will be described
elsewhere) shows that there is a symmetry restoring (second order) phase
transition both at high temperature and high density. The critical temperature
$T_c$ is $T_c\sim m_{dyn}$, and the critical density $n_c$ is
$n_c=\frac{|eB|}{2\pi}$ at $g\leq g_c=\sqrt{\pi}$ (corresponding to the
filling of the lowest fermion Landau level) and $n_c>\frac{|eB|}{2\pi}$ at
$g>g_c$. We recall that there cannot be spontaneous breakdown of a continuous
symmetry at finite $(T>0)$ temperature in $2+1$ dimensions (the
Mermin-Wagner-Coleman (MWC) theorem [11,12]). In the NJL model, the MWC
theorem manifests itself only beyond the leading order in $1/N_c$. A plausible
hypothesis, we believe, is that beyond this order, the phase transition at
$T=T_c\sim m_{dyn}$ will become the Berezinsky-Kosterlitz-Thouless (BKT) type
phase transition [13], and, in particular, at $0<T<T_c$, the NG modes
$\tau$ and $\pi$ will transform into BKT collective modes.

In this Letter, we considered the dynamics in the presence of a constant
magnetic field only. It would be interesting to extend this analysis to the
case of inhomogeneous electromagnetic fields in $2+1$ dimensions. As we have
been informed recently, the program of the derivation of a low energy effective
action in $2+1$ dimensional QED in external electromagnetic fields is now being
developed by ~Cangemi, ~D'Hoker, and ~Dunne [14].

V.A.M. is grateful to the members of the Department of Applied Mathematics of
the University of Western Ontario, where part of this  work was done,
and of the Institute for Theoretical Physics of the University of California
(Santa Barbara) for their hospitality. He thanks J.M.~Cornwall, E.~D'Hoker,
D.~Kaplan,       J.~Polchinski,
S.~Raby, L.~Randall, J.~Schwarz,
 A.~Vainshtein,   K.~Yamawaki,
and A.~Zee for useful discussions.

The research was supported in part by the National Science Foundation under
Grant No. PHY89-04035. I.A.S. is grateful to the Soros Science Foundation for
financial support.

\end{document}